\documentclass[12pt,oneside]{article}
\newcommand{\be}{\begin{equation}}
\newcommand{\ee}{\end{equation}}
\newcommand{\Tr}{{\rm Tr}}
\def\bea{\begin{eqnarray}}
\def\eea{\end{eqnarray}}
\def\bean{\begin{eqnarray*}}
\def\eean{\end{eqnarray*}}
\def\thru#1{\mathrel{\mathop{#1\!\!\!\!/}}}

\def\part#1{\partial^{}_#1}

\begin{document}
\thispagestyle{empty}
\setcounter{page}{0}
\renewcommand{\theequation}{\thesection.\arabic{equation}}

{\hfill{UFIFT-HET-01-27}} 

%{\hfill{\tt hep-th/0112261}}

\vspace{2cm}

\begin{center}
{\bf ALGEBRAIC DREAMS}

\vspace{1.4cm}

PIERRE RAMOND

\vspace{.2cm}

{\em Institute for Fundamental Theory,} \\
{\em Department of Physics, University of Florida } \\
{\em Gainesville FL 32611, USA}
\end{center}

%\centerline{\large\bf  ALGEBRAIC DREAMS}
%\vskip 1cm
%\centerline{\bf Pierre Ramond} 
%\vskip .5cm
%\centerline{\em  Institute for Fundamental Theory,}
%\centerline{\em Department of Physics, University of Florida}
%\centerline{\em Gainesville FL 32611, USA}
%\vskip 1.5cm
\vspace{-.1cm}

\centerline{{\tt ramond@phys.ufl.edu}}

\vspace{1cm}

\centerline{ABSTRACT}

\vspace{- 4 mm}  

\begin{quote}\small
Nature's attraction to unique mathematical structures provides powerful 
hints for unraveling her mysteries. None is at present as intriguing as eleven-dimensional M-theory. The search for exceptional structures specific to eleven dimensions  leads us to exceptional groups in the description of space-time. One specific connection, through the coset $F_4/SO(9)$, may provide  a generalization of eleven-dimensional supergravity. Since this coset happens to be the projective space of the Exceptional Jordan Algebra, its charge space may be linked to the fundamental degrees of freedom underlying M-theory. 
\end{quote}

\baselineskip18pt

\newpage

\setcounter{equation}{0}
\section{Introduction}
  Nature relishes unique mathematical structures. A prime example is the consistency of superstring theory which seems, time and again, to rely on miracles often traced to algebraic ``coincidences" that naturally appear   in special algebraic  structures. This  provides us mortals  with a strong guiding principle to unravel Her mysteries.

Nature shows that space-time symmetries with dynamics associated with gravity, and internal symmetries with their dynamics described by Yang-Mills theories, can coexist peacefully.    How does She do it? M-theory and Superstring theories~\cite{JOHN} are the only examples of theories where this union appears possible, but there remain important unanswered questions. While  theory is formulated in  terms of local operators in space-time,  space-time as we know it, is probably not as fundamental as we think, and only a solution of a more general theory. Does this solution contain clues  as to the nature of the underlying theory? In Quantum Mechanics, time is treated very differently from space, and space-time symmetries do not seem natural, leading  some   to speculate~\cite{THORN} that only compact 
symmetries reflect those of the underlying theory. This line of reasoning, applied to 
supersymmetry, leads to all types of questions: is it clear that M-theory is manifestly supersymmetric in eleven dimensions? Could it be supersymmetric only in the local limit or when compactified to lower dimensions?

\section{Is Space-Time Exceptional?}
 The Exceptional Algebras are   most unique and beautiful among Lie Algebras, and no one should be surprised if Nature uses them. To that effect, we present some mathematical and physical factoids which may suggest new lines of exceptional inquiry:

\begin{itemize}

\item Patterns of the quantum numbers of the elementary particles point to their embedding in Exceptional Groups~\cite{ESIX}. The sequence 

$$
E^{}_8~\supset~ E^{}_7~\supset E^{}_6~\supset~SO(10)~\supset~SU(5) ~\supset SU(3)\times SU(2) \ ,$$
obtained by chopping off one dot from these algebras' Dynkin diagrams leads to the  non-Abelian symmetries of the Standard Model. This sequence is realized in the Heterotic string~\cite{HET} where the gauge algebra is $E_8\times E_8$.

\item Exceptional Groups contain orthogonal groups capable of describing space-time symmetries. Some compact embeddings are 

$$E^{}_8\,\supset\, SO(16)\ ,\qquad E^{}_7\,\supset\, SO(12)\times SO(3)\ ,\qquad E^{}_6\,\supset\,SO(10)\times SO(2)\ .$$
This may occur along non-compact groups as well, for instance along the  sequence from $E_{8\,(-24)}$ to $SO(9)$:   

$$E^{}_{8\,(-24)}\,\supset\,E^{}_{7\,(-25)}\times SU(1,1)\ ,\qquad  E^{}_{7\,(-25)}\,\supset\,E_{6\,(-26)}^{}\times SO(1,1)\ ,$$
$$
E_{6\,(-26)}^{}\,\supset\,F_{4\,(-20)}^{}\,\supset\,SO(9)\ .$$
 
\item The consistency of superstring~\cite{R} theories in $9+1$ dimensions  relies on  the triality of the light-cone little group $SO(8)$,  which links its tensor and spinor representations via a $Z_3$ symmetry. The exceptional group $F_4$ is the smallest which realizes this triality explicitly. It was surprising to find another consistent theory in one more space dimension since the $SO(9)$ little group has very different spinor and tensor representations. A possible hint for fermion-boson confusion is the anomalous embedding of $SO(9)$ into an orthogonal group in which the vector representation of the bigger group is identified with the spinor of the smaller group

$$
SO(16)\supset SO(9)\ ,\qquad {\bf 16}~=~{\bf 16}\ .$$

\item The use of exceptional groups to describe space-time symmetries has not been as fruitful. One obstacle has been that exceptional algebras relate tensor and spinor representations of their orthogonal subgroups, while Spin-Statistics requires them to be treated differently. Yet there are some  mathematical curiosities worth noting. For one, the anomalous Dynkin embedding of $F_4$ inside $SO(26)$

$$
SO(26)\supset F^{}_4\ ,\qquad  {\bf 26}~=~{\bf 26}\ ,$$
or its non-compact variety

$$SO(25,1)\,\supset\,F_{4\,(-20)}^{}\ ,$$
together with the embedding

$$
F^{}_4~\supset~SO(9)\ ,\qquad {\bf 26}~=~{\bf 1}\,\oplus\,{\bf 9}\,\oplus\,{\bf 16}\ ,$$
or its non-compact form $F_{4\,(-20)}^{}\,\supset\,SO(9)$, might point to a (M)-heterotic construction from the bosonic string to M-theory.

\item A  formulation of finite-dimensional Hilbert spaces in terms of the algebra of observables, proposed by P. Jordan~\cite{JORDAN}, has not yet proven fruitful in Physics, in spite of many attempts. In all but one case, it is akin to rewriting the familiar Dirac ket description in terms of density matrices, but it also unearthed a unique structure on which Quantum Mechanics can be implemented, even though it cannot be described by kets in Hilbert space. Our interest lies in the fact that its automorphism group is $F_4$ and its natural description lies in the sixteen-dimensional (Cayley) projective space $F_4/SO(9)$. 

\item 
There is a whiff of the exceptional group $F_4^{}$ in the supergravity supermultiplet in eleven dimensions, as we now proceed to show.   

\end{itemize}

\section{Supergravity in Eleven Dimensions}
Eleven dimensional $N=1$ Supergravity~\cite{ELEVEN}  is the ultimate field theory that includes gravity, but it is not renormalizable, and does not stand on its own as a physical theory. 
Its counterpart, the ultimate field theory without gravity, is the finite $N=4$ Super Yang-Mills theory in four dimensions. Recently, the  eleven-dimensional theory has been revived as the limit of M-theory which, like characters on the walls of Plato's  cave, has revealed itself through its compactified version onto lower-dimensional manifolds. In the absence of  a definitive description of M-theory, it behooves us to scrutinize what is known about $11$-d Supergravity theory. 

\subsection{SuperAlgebra}
$N=1$ supergravity in eleven dimension is a local field theory that contains three  massless fields, the familiar symmetric second-rank tensor, $h^{}_{ \mu\nu}$ which represents gravity,  a three-form field $A^{}_{\mu\nu\rho}$, and the Rarita-Schwinger spinor $\Psi^{}_{\mu\,\alpha}$.  From its Lagrangian, one can derive the expression for the super Poincar\'e algebra, which in the unitary transverse gauge assumes the particularly simple form in terms of the nine $(16\times 16)$ $\gamma_i$ matrices which form the Clifford algebra

$$\{\,\gamma_{}^i,\gamma_{}^j\,\}~=~2\delta^{ij}_{}\ ,~~~~~~i,j~=1,\dots,9\ .$$
Supersymmetry is generated by the sixteen real supercharges 
$$
{\cal Q}^{a}_{\pm}={\cal Q}_{\pm}^{a\,*}\ ,
\nonumber$$
 which satisfy 
$$
\{{\cal Q}^a_+,{\cal Q}^b_+\}\,=\,\sqrt{2}\,p^+\delta^{ab}\ ,\qquad 
\{{\cal Q}^a_-,{\cal Q}^b_-\}\,=\,\frac{{\vec p}\cdot{\vec p}}{\sqrt{2}\,p^+}\delta^{ab}\ ,\qquad
\{{\cal Q}^a_+,{\cal Q}^b_-\}\,=\,-(\gamma_i)^{ab}p^i\ ,
$$
and  transform  as Lorentz spinors
\bean
~[M^{ij},{\cal Q}^a_\pm]&=&\frac i2(\gamma_{}^{ij}{\cal Q}_\pm)^a\ ,\qquad 
~[M^{+-},{\cal Q}^a_\pm]~=~\pm\frac i2{\cal Q}_\pm^a\ ,\\
~[M^{\pm i},{\cal Q}^a_\mp]&=&0\ ,\qquad 
~[M^{\pm i},{\cal Q}^a_\mp]~=~\pm\frac i{\sqrt{2}}(\gamma_{}^i{\cal Q}_\pm)^a\ .
\eean
A very simple representation of the $11$-dimensional super-Poincar\'e generators can be constructed,
in terms of sixteen anticommuting real $\chi$'s and their derivatives, which transform as the spinor of $SO(9)$, as
$$
{\cal Q}_+^{a}~=~\partial_{\chi^a}+\frac{1}{\sqrt{2}}p^+\chi^a\ ,\qquad {\cal Q}_-^a~=~
-\frac{p_{}^i}{p^+}\;\left(\gamma_{}^i\,{\cal Q}_+\right)^a_{}\ ,$$

\bean
~M^{ij}&=&x^ip^j-x^jp^i-\frac{i}{2}\chi\;\gamma_{}^{ij}\partial_\chi\ ,\\
~M^{+-}&=&-x^-p^+ -\frac{i}{2}\chi\;\partial_\chi \ ,\\
~M^{+i}&=&-x^ip^+\ ,\\
~M^{-i}&=&x^-p^i-\frac{1}{2}\{x^i,P^-\}+\frac {ip^j}{2p^+}\chi\gamma_{}^i\gamma_{}^j\partial_\chi  \ .
\eean
The light-cone little group transformations are  generated by   

$$
S^{ij}_{}~=~-\frac i2\chi\,\gamma^{ij}_{}\,\partial_\chi\ ,$$
which satisfy the  $SO(9)$ Lie algebra. In order to examine the spectrum,  we  rewrite the supercharges in terms of eight complex Grassmann variables

$$
\theta^\alpha~\equiv~\frac{1}{\sqrt{2}}\left(\chi^\alpha+i\chi^{\alpha+8}\right)\ ,\qquad \overline\theta^\alpha ~\equiv~\frac{1}{\sqrt{2}}\left(\chi^\alpha-i\chi^{\alpha+8}\right)\ ,\label{choice}$$
and 

$$
\frac{\partial}{\partial\theta^\alpha}~\equiv~\frac{1}{\sqrt{2}}\left(\frac{\partial}{\partial\chi^\alpha}-i
\frac{\partial}{\partial\chi^{\alpha+8}}\right)\ ,\qquad 
\frac{\partial}{\partial\overline\theta^\alpha}~\equiv~
\frac{1}{\sqrt{2}}\left(\frac{\partial}{\partial\chi^\alpha}+i\frac{\partial}{\partial\chi^{\alpha+8}}\right)\ ,$$
where $\alpha=1,2,\dots,8$. The eight complex $\theta$  transform as the $({\bf 4}\, ,\, {\bf 2})$,   and $\overline\theta$ as the $(\overline{\bf 4}\,,\, {\bf 2}) $ of the $SU(4)\times SU(2)$ subgroup of $SO(9)$.  The eight complex supercharges 

\bean
{\bf Q}_+^\alpha
&\equiv &\frac{1}{\sqrt{2}}\left({\cal Q}_+^\alpha +i{\cal Q}_+^{\alpha+8}\right)~=~\frac{\partial}{\partial\overline\theta^\alpha}+
\frac{1}{\sqrt{2}}p^+\theta^\alpha\ ,\\
 {\bf Q}_+^{\alpha\,\dagger}
&\equiv&\frac{1}{\sqrt{2}}\left({\cal Q}_+^\alpha -i{\cal Q}_+^{\alpha+8}\right)~=~\frac{\partial}{\partial\theta^\alpha}+\frac{1}{\sqrt{2}}p^+\overline\theta^\alpha\ ,
\eean 
 satisfy

$$
\{\,{\bf Q}_+^{\alpha}\,,\,{\bf Q}_+^{\beta\,\dagger}\,\}~=~\sqrt{2}\,p^+\,\delta^{\alpha\beta}\ .
$$
They act irreducibly on  chiral superfields  which are annihilated by the covariant derivatives

$$
\left(\frac{\partial}{\partial\overline\theta^\alpha}-
\frac{1}{\sqrt{2}}p^+\theta^\alpha\right)\,\Phi(y^-,\theta)~=~0\ ,$$
where

$$
y^-~=~x^--\frac{i\theta\overline\theta}{\sqrt{2}}\ .$$
Expansion of the superfield in powers of the eight complex $\theta$'s yields $256$ components, with the following $SU(4)\times SU(2)$ properties
\bean
1~~~&\sim&~~~({\bf 1},{\bf 1})\ ,\\
\theta~~~&\sim&~~~({\bf 4},{\bf 2})\ ,\\
\theta\theta~~~&\sim&~~~({\bf 6},{\bf 3})\oplus({\bf 10},{\bf 1})\ ,\\
\theta\theta\theta~~~&\sim&~~~({\bf \overline {20}},{\bf 2})\oplus({\bf \overline {4} },{\bf  4})\ ,\\
\theta\theta\theta\theta~~~&\sim&~~~({\bf 15 },{\bf 3 })\oplus({\bf 1 },{\bf  5})\oplus({\bf 20' },{\bf 1 })\ ,\eean
and the higher powers yield the conjugate representations by duality. These   make up the three representations of $N=1$ supergravity

\bean
{\bf 44}~&=&~({\bf 1  },{\bf 5 })\oplus({\bf 6 },{\bf  3})\oplus({\bf 20' },{\bf  1})\oplus({\bf 1 },{\bf  1})\ ,\\
{\bf 84}~&=&~({\bf 15  },{\bf 3 })\oplus({\bf \overline {10} },{\bf  1})\oplus({\bf 10 },{\bf  1})\oplus({\bf 6 },{\bf  3})\oplus({\bf 1 },{\bf  1})\ ,\\
{\bf 128}~&=&~({\bf 20  },{\bf 2 })\oplus({\bf \overline {20} },{\bf  2})\oplus({\bf 4 },{\bf  4})\oplus({\bf \overline {4} },{\bf  4})\oplus({\bf 4 },{\bf  2})\oplus({\bf \overline {4} },{\bf  2})\ .\eean
For future reference we note the $SU(4)\times SU(2)$ weights of the $\theta$s, using the notation $(a_1,a_2,a_3;\,a)$,  

\bean
\theta^1&\sim(1,0,0;\, 1)\ ,&\theta^8\sim(1,0,0;\, -1)\ ,\\ 
\theta^4&\sim(-1,1,0;\, 1)\ ,&\theta^5\sim(-1,1,0;\, -1)\ ,\\
\theta^7&\sim(0,-1,1;\, 1)\ ,&\theta^2\sim(0,-1,1;\, -1)\ ,\\ 
\theta^6&\sim(0,0,-1;\, 1)\ ,&\theta^3\sim(0,0,-1;\, -1)\ , 
\eean 
which enables  us to find the  highest weights of the supergravity representations

\bean {\bf 44}~~&:&~~~\theta^1\theta^4\theta^5\theta^8~=~(0,2,0;\,0)~~\sim~~({\bf 20'}\,\,{\bf 1})\\
{\bf 84}~~&:&~~~\theta^1\theta^8~~~~~~~=~(2,0,0;\,0)~~\sim~~({\bf 10}\,,{\bf 1}\,)\\
{\bf 128}~~&:&~~~\theta^1\theta^4\theta^8~~~~=~(1,1,0;\,1)~~\sim~~({\bf 20}\,,{\bf 2}\,)\ ,\eean
together with their $SU(4)\times SU(2)$ properties. All other states are generated by acting on these highest weight states with the lowering operators. The highest weight chiral superfield that describes $N=1$ supergravity in eleven dimensions is simply

$$
\Phi~=~\theta^1\theta^8\,h(y^-,\vec x)~+~\theta^1\theta^4\theta^8\,\psi(y^-,\vec x)~+~\theta^1\theta^4\theta^5\theta^8\,A(y^-,\vec x)\ ,$$
which summarizes the spectrum of the super-Poincar\'e algebra in eleven dimensions of either a free  field theory or a free superparticle.
\vskip 1cm

Since the little group generators  act on a $256$-dimensional space, we can express them in terms  of sixteen $(256\times 256)$ matrices, $\Gamma^a_{}$, which   satisfy the Dirac algebra

$$\{\, \Gamma_{}^a\,,\, \Gamma_{}^b\,\}~=~\,2\delta_{}^{ab}\ .$$ 
 This leads to an  elegant representation of the $SO(9)$ generators

$$
S^{ij}=-\frac {i}{4}(\gamma^{ij})^{ab}\, \Gamma_{}^a\, \Gamma_{}^b~\equiv~-\frac{i}{2}f^{\,ij\,a\,b} \Gamma_{}^a\, \Gamma_{}^b\ . \label{S_ij} $$
The coefficients

$$ f^{\,ij\,a\,b}~\equiv~ \frac{1}{2}(\gamma_{}^{ij})^{ab}\ ,$$
naturally appear in the commutator between the generators of $SO(9)$ and any spinor operator  $T_{}^a$, as 

$$[\,T_{}^{ij}\,,\,T_{}^a\,]~=~\frac{i}{2}\left(\gamma_{}^{ij}\,T\right)^a~=~if^{\,ij\,a\,b}\,T_{}^b\ .$$
But there is more to it, the $(\gamma_{}^{ij})^{ab}$ can also be viewed as structure constants of a Lie algebra. Manifestly antisymmetric  under $a\leftrightarrow b$,  they can appear  in the commutator of two spinors into the $SO(9)$ generators   

$$ [\,T_{}^a\,,\,T_{}^b\,]~=~ \frac {i}{2}(\gamma^{ij})^{ab} \,T_{}^{ij} ~=~\,f^{\,a\,b\,ij}\,T_{}^{ij}\ ,$$
and one easily checks that  they satisfy the Jacobi identities. Remarkably, the $52$ operators  $T^{ij}$ and $T^a$ generate the exceptional Lie algebra $F_4$, showing explicitly how an exceptional Lie algebra appears in the light-cone formulation of supergravity in eleven dimensions! 

\subsection{Character Formula}
The   degrees of freedom are labelled by the light-cone little group  $SO(9)$ acting on the  transverse vector indices, as $h_{(ij)}\sim (2000)$, $A_{[ijk]}\sim (0010)$, $\Psi_{i\,\alpha}\sim (1001)$, with their little group representations in Dynkin's notation. 
 
Their group-theoretical properties are summarized in the following table 
\hskip 2cm
\begin{center}
\begin{tabular}{|c|c|c|c|}
\hline
$~{\rm irrep}~$& $(1001)$&$ (2000)$ & $(0010)$   \\
 \hline \hline         
$~D~ $&$  128$ & $ 44$ & $ 84$  \\
 \hline    
$~I_2~$& $256$& $88$ & $168$  \\
\hline
$~I_4~$& $640$& $232$ & $408$ \\                                                            
 \hline
$~I_6~$&$1792$& $712$ &$1080$\\ 
\hline
$~I_8~$&$5248$& $2440$ &$3000$\\ 
\hline
 \end{tabular}\end{center}
\vskip 0.3cm
where $D$ is the dimension of the representation, and $I^{}_n$ are the Dynkin indices of the representations, related to the four  Casimir operators of $SO(9)$. We note that the dimension and Dynkin indices of the fermion is the sum over those of the bosons, except for $I_8$, indicating that these three representations have much in common.   

Amazingly, the supergravity fields  are the  first of an infinite number of triplets~\cite{PR} of $SO(9)$ representations which display the same group-theoretical relations: equality of dimension and all Dynkin indices except $I_8$ between one representation and the sum of the other two. Quantum theories of these  Euler triplets may  have very interesting divergence properties, as these numbers typically occur in higher loop calculations, and such equalities usually increase the degree of divergence, and the failure of the equality for $I_8$ is probably related to the lack of renormalizability of the theory~\cite{CURTRIGHT}. 

This mathematical fact has been traced to a  character formula~\cite{GKRS} related to the three equivalent embeddings of $SO(9)$ into $F_4$! The character formula is given by  

$$
V_{\lambda}\,\otimes\,S^+_{}\,-\,V_{\lambda}\,\otimes\,S^-_{}~=~\sum_{c }\,{\rm sgn}(c)\,U_{c\bullet\lambda}\ .$$
On the left-hand side,  $V_\lambda$ is a representation of $F_4$ written in terms of its $SO(9)$ subgroup, $S^\pm$ are the two spinor representations of $SO(16)$ written in terms of its anomalously embedded subgroup $SO(9)$, $\otimes$ denotes the normal Kronecker product of representations, and the $-$ denotes the naive substraction of representations. On the right-hand side, the sum is over  $c$, the  elements of the Weyl group  which map the Weyl chamber of $F_4$ into that of $SO(9)$.  In this case there are three elements, the ratio of the orders of the Weyl groups (it is also the Euler number of the coset manifold), and $U_{c\bullet\lambda}$ denotes the $SO(9)$ representation with highest weight $c\bullet\lambda$,  where

$$c\bullet\lambda~=~c\,(\lambda+\rho^{{}^{}}_{F_4})-\rho^{{}^{}}_{SO(9)}\ ,$$ 
and the $\rho$'s are the sum of the fundamental weights for each group, and ${\rm sgn}(c)$ is the index of $c$. Thus to each $F_4$ representation corresponds a triplet, called Euler triplet. The supergravity case is rather trivial as 

$$SO(16)\,\supset\,SO(9)\ ,\qquad S^+_{}\,\sim\,{\bf 128}={\bf 128}\ ,\qquad S^-_{}\,\sim\,{\bf 128}'={\bf 44}\,+\, {\bf 84}\ ,$$
and the character formula reduces to the  truism

$${\bf 128}\,-\,{\bf 44}\,-\, {\bf 84}~=~{\bf 128}\,-\,{\bf 44}\,-\, {\bf 84}\ .$$
This construction yields the general form of the Euler triplets: the Euler triplet corresponding to the $F_4$ representation $[\,a_1\,a_2\,a_3\,a_4\,]$ is made up of the following three $SO(9)$ representations listed in order of increasing dimensions:

$$
 (2+a_2+a_3+a_4,a_1,a_2,a_3)\ ,~ (a_2,a_1,1+a_2+a_3,a_4)\ ,~(1+a_2+a_3,a_1,a_2,1+a_3+a_4)$$
The spinor representations appear with odd entries in the fourth place. Euler triplets with the largest spinor and two bosons must have both $a_3$ and $a_4$ even or zero. 

Since the Dynkin indices of the product of two representations satisfy the composition law

$$I_{}^{(n)}[\lambda\otimes \mu]~=~d^{}_\lambda\,I_{}^{(n)}[\mu]+d_\mu^{}\,I_{}^{(n)}[\lambda]\ ,$$
it follows that the deficit in $I^{(8)}$ is always proportional to the dimension of the $F_4$ representation that generates it. 
\subsection{The Kostant Operator}
This character formula can be viewed as the  index formula of a  Dirac-like operator~\cite{KOS} formed over the  coset $F^{}_4/SO(9)$. This coset is the sixteen-dimensional Cayley projective plane, over which we introduce the previously considered Clifford algebra    

$$
\{\, \Gamma_{}^a\,,\, \Gamma_{}^b\,\}~=~2\,\delta_{}^{ab}\ ,~~a,b=1,2,\dots, 16\ ,$$
generated by $(256\times 256)$ matrices. The Kostant equation is defined as 

$$
\thru {\cal K}\,\Psi~=~\sum_{a=1}^{16}\, \Gamma_{}^a\,T^{a}_{}\,\Psi~=~0\ ,$$
where $T_a$ are $F_4$ generators not in $SO(9)$, with  commutation relations

$$
[\,T_{}^a\,,\,T_{}^b\,]~=~i\,f^{\,ab\,ij}_{}\,T^{ij}_{}\ .$$
Although it is taken over a compact manifold, it has non-trivial solutions. To see this, we rewrite its square as the difference of positive definite quantities, 

$$
\thru {\cal K}\, \thru {\cal K}\,~=~C^{2}_{F_4}-C^{2}_{SO(9)}+72\ ,$$
where 

$$
C^{2}_{F_4}~=~\frac 12\,T_{}^{ij}\,T_{}^{ij}+T_{}^a\,T_{}^a\ ,$$
is the $F_4$ quadratic Casimir operator, and

$$
C^{2}_{SO(9)}~=~\frac 12\,\left(T_{}^{ij}-if_{}^{ab\,ij}\,\widetilde\Gamma_{}^{ab}\right)^2\ ,$$
 is the quadratic Casimir for the sum

$$
L_{}^{ij}~\equiv~T_{}^{ij}+S_{}^{ij}\ ,$$
where $S^{ij}$ is the previously defined $SO(9)$ generator which acts on the supergravity fileds.  We have also used  the quadratic Casimir on the spinor representation

$$
 \frac 12\,S_{}^{ij}\,S_{}^{ij}~=~72\ .$$
Kostant's operator commutes with the sum of the generators,

$$
[\,\thru {\cal K}\,,\,L_{}^{ij}\,]~=~0\ ,$$
allowing  its solutions to be labelled by $SO(9)$ quantum numbers. 

The same construction of Kostant's operator applies to all  equal rank embeddings, and its trivial solutions display supersymmetry~\cite{GKRS,BR,BR2,PR2}. In particular we note the cases $E_6/SO(10)\times SO(2)$, with Euler number $27$, $E_7/SO(12)\times SO(3)$ with Euler number $63$, and $E_8/SO(16)$, where the Euler triplets contain $135$ representations~\cite{PR}. These cosets with dimensions $32\ ,64$, and $128$ could be viewed as complex, quaternionic and octonionic Cayley plane~\cite{ATIYAH}

\subsection{Oscillator Representation of $F_4$ }
Schwinger's celebrated representation of $SU(2)$  generators of in terms of one doublet of harmonic oscillators can be extended to other Lie algebras~\cite{FULTON}. The generalization involves several sets of harmonic oscillators, each spanning  the fundamental representation. For example, $SU(3)$ is generated by two sets of  triplet harmonic oscillators, $SU(4)$   by two quartets. In the same way, all representations of the exceptional group $F_4$ are generated by three sets of oscillators transforming as  ${\bf 26}$.  We label each copy of $26$ oscillators
as $A^{[\kappa]}_0,\; A^{[\kappa]}_i,\; i=1,\cdots,9,\; B^{[\kappa]}_a,\; a=1,\cdots,16$, 
and their hermitian conjugates, and where $\kappa=1,2,3 $. 
Under $SO(9)$, the $A^{[\kappa]}_i$ transform as ${\bf 9}$, $B^{[\kappa]}_a$ transform as ${\bf 16}$, 
and $A^{[\kappa]}_0$ is a scalar. They satisfy the commutation relations of ordinary   harmonic oscillators

$$
[\,A^{[\kappa]}_i\,,\,A^{[\kappa']\,\dagger}_j\, ]~=~\delta^{}_{ij}\,\delta_{}^{[\kappa]\,[\kappa']}\ ,\qquad [\,A^{[\kappa]}_0\,,\,A^{[\kappa']\,\dagger}_0\, ]~=~\delta_{}^{[\kappa\,\kappa']}\ . 
$$
Note that the  $SO(9)$  spinor operators   satisfy Bose-like commutation relations

$$ \nonumber[\,B^{[\kappa]}_a\,,\,B^{[\kappa']\,\dagger}_b\, ]~=~\delta^{}_{ab}\,\delta_{}^{[\kappa]\,[\kappa']}\ .$$
The   generators $T_{ij}$ and $T_a$  

\begin{eqnarray}
T^{}_{ij}&=&-i\sum_{\kappa=1}^4\left\{\left(A^{[\kappa]\dag}_iA^{[\kappa]}_j-A^{[\kappa]\dag]}_jA^{[\kappa]}_i\right)+\frac 12\,B^{[\kappa]\dag}\,\gamma^{}_{ij} B^{[\kappa]}\label{t_ij}\right\}\nonumber\ ,\\
T_a&=&-\frac{i}{{2}}\sum_{\kappa=1}^4\left\{ (\gamma_i)^{ab}\left(A^{[\kappa]\dag}_iB^{[\kappa]}_b-B^{[\kappa]\dag}_bA^{[\kappa]}_i\right)-\sqrt{3}\left(B^{[\kappa]\dag}_aA^{[\kappa]}_0-A^{[\kappa]\dag}_0B^{[\kappa]}_a\right)\right\}\nonumber\ ,\label{t_a}
\end{eqnarray}
satisfy the $F_4$ algebra,

\bean
[\,T^{}_{ij}\,,T^{}_{kl}\,]&=&-i\,(\delta^{}_{jk}\,T^{}_{il}+\delta^{}_{il}\,T^{}_{jk}-\delta^{}_{ik}\,T^{}_{jl}-\delta^{}_{jl}\,T^{}_{ik})\ ,\\
~[\,T^{}_{ij}\,,T^{}_a\,]&=&\frac i2\,(\gamma^{}_{ij})^{}_{ab}\,T^{}_b\ ,\\
~[\,T^{}_a\,,T^{}_b\,]&=&\frac i2\,(\gamma^{}_{ij})^{}_{ab}\,T^{}_{ij}\ ,
\eean
so that the structure constants are given by

$$
f_{ij\,ab}~=~f_{ab\,ij}~=~\frac{1}{2}\,(\gamma^{}_{ij})_{ab}\ .$$
The last commutator  requires the Fierz-derived identity  

$$
\frac{1}{4}\,\theta\,\gamma^{ij}\,\theta\;\chi\,\gamma^{ij}\,\chi~=~3\,\theta\,\chi\;\chi\,\theta+\theta\,\gamma^i\,\chi\;\chi\,\gamma^i\theta\ ,$$
from which we deduce

$$
3\,\delta^{ac}\delta^{db}+(\gamma^i)^{ac}\,(\gamma^i)^{db}- (a\leftrightarrow b)~=~
\frac{1}{4}\,(\gamma^{ij})^{ab}\,(\gamma^{ij})^{cd}\ .$$
To satisfy these commutation relations, we have required  both $A_0$ and $B_a$ to obey Bose commutation relations (Curiously, if  both  are anticommuting, the $F_4$ algebra is still satisfied). One can just as easily use a coordinate representation of the oscillators by introducing real coordinates $u^{}_i$ which transform as transverse space vectors, $u^{}_0$ as scalars, and $\zeta^{}_a $ as space spinors which satisfy Bose commutation rules

\bean
A_i&=&\frac 1{\sqrt{2}}(u^{}_i+\partial_{u^{}_i})\ ,\qquad A^\dag_i=\frac 1{\sqrt{2}}(u^{}_i-\partial_{u^{}_i})\ ,\\
B_a&=&\frac 1{\sqrt{2}}(\zeta^{}_a+\partial_{\zeta^{}_a})\ ,\qquad B^\dag_a=\frac 1{\sqrt{2}}(\zeta^{}_a-\partial_{\zeta^{}_a})\ ,\\
A_0&=&\frac 1{\sqrt{2}}(u^{}_0+\partial_{u^{}_0})\ ,\qquad A^\dag_0=\frac 1{\sqrt{2}}(u^{}_0-\partial_{u^{}_0})\ .
\eean
Using square brackets $[\cdots]$ to represent the Dynkin label of $F_4$, and
round brackets $(\cdots)$ to represent those of $SO(9)$, we list some of the combinations which will be used for investigating the solutions of Kostant's equation

\bean
u^{}_1+iu^{}_2&\sim& [\,0~~\,0~~\,0~~\,1\,]~\sim~ (~\,1~~\,0~~\,0~~\,0~)~ \ , \\
u^{}_3+iu^{}_4&\sim& [\,1~~\,0~~\,0-\!\! 1\,]~\sim~ (-\!\! 1~~\,1~~\,0~~\,0~) \ , \\
\zeta^{}_1+i\zeta^{}_9&\sim &[\,0~~\,0~~\,1-\!\!1\,]~\sim~ (~\,0~~\,0~~\,0~~\,1~) \ , \\
\zeta^{}_8+i\zeta^{}_{16}&\sim& [\,0~~\,1-\!\!1~~\,0\,]~\sim ~(~\,0~~\,0~~\,1-\!\! 1~) \ , \\
\zeta^{}_3-i\zeta^{}_{11}&\sim& [\,1-\!\!1~~\,1~~\,0\,]~\sim~ (~\,0~~\,1-\!\! 1~~\,1~) \ , \\
\zeta^{}_6-i\zeta^{}_{14}&\sim& [\,1~~\,0-\!\!1~~\,1\,]~\sim~ (~\,0~~\,1~~\,0-\!\! 1~)  \ . 
\eean
Hence $u^{}_1+iu^{}_2$ and $\zeta^{}_1+i\zeta^{}_9$  are the highest weights of the $SO(9)$ representations $\bf 9$, and $\bf 16$, respectively.

\subsection{Solutions of Kostant's Equation}
For every representation of $F_4$,  $[a_1,a_2,a_3,a_4]$, there is one $SO(9)$ Euler triplet solution of Kostant's equation 

$$
 (2+a_2+a_3+a_4,a_1,a_2,a_3)\ ,~~(a_2,a_1,1+a_2+a_3,a_4)\ ,~~(1+a_2+a_3,a_1,a_2,1+a_3+a_4)
$$
The trivial solution with   $a_1=a_2=a_3=a_4=0$,  yields the  $N=1$ supergravity multiplet in eleven dimensions, $(2000)\oplus(0010)\oplus(1001)$. We have seen that the highest weight solution are $\theta^1\theta^4\theta^5\theta^8\ , \theta^1\theta^8$, and $\theta^1\theta^4\theta^8$, described by the chiral superfield 

$$
\Phi_{0000}^{}~=~\theta^1\theta^8\,h_{0000}^{}(y^-,\vec x)~+~\theta^1\theta^4\theta^8\,\psi_{0000}^{}(y^-,\vec x)~+~\theta^1\theta^4\theta^5\theta^8\,A_{0000}^{}(y^-,\vec x)\ .$$
In general, the  highest weight solutions appear in the form of $f(u_i,\zeta_a)\,\Theta(\theta)$,   where both $f(u_i,\zeta_a)$ and $\Theta(\theta)$ are the highest weights  $SO(9)$ states  with respect to the earlier defined  $T_{ij}$ and $S_{ij}$. The solutions have the quantum numbers of their sum  $L_{ij}=S_{ij}+T_{ij}$, which commutes with Kostant's operator.    $\Theta(\theta)$ is one of the three polynomials above, $\theta^1\theta^4\theta^5\theta^8\ , \theta^1\theta^8$, or $\theta^1\theta^4\theta^8$.

The highest weight solutions corresponding to each fundamental representation of $F_4$ are~\cite{NEXT}

\begin{enumerate} 

\item{$a_1=a_2=a_3=0,\, a_4\ge 1$} 

\noindent These representations are built with only one copy. The  highest weight solutions with $a_4=1$ are uniquely given by

\bean
&& \theta^1\theta^4\theta^5\theta^8(u^{}_1+iu^{}_2), \\ &&\theta^1\theta^8(\zeta^{}_1+i\zeta^{}_9)\ ,\nonumber\\ 
&&\theta^1\theta^4\theta^8(\zeta^{}_1+i\zeta^{}_9)\ ,\nonumber
\eean
where $(u^{}_1+iu^{}_2)$ and $(\zeta^{}_1+i\zeta^{}_9)$ are the highest weights of $SO(9)$ representations $(1000)$ and $(0001)$, respectively.

\item{$a_1=a_2=a_4=0,\, a_3\ge 1$}

\noindent In this case we need two copies ($\kappa=1,2$). For $a_3=1$, the highest weight 
solutions are

\begin{eqnarray}
&&\theta^1\theta^4\theta^5\theta^8\,[\,u_1+iu_2\,,\,\zeta_1+i\zeta_9\,]\ ,\nonumber\\
&&\theta^1\theta^8\,[\,\zeta_1+i\zeta_9\,,\,\zeta_8+i\zeta_{16}\,]\ ,\nonumber\\
&&\theta^1\theta^4\theta^8\,[\,u_1+iu_2\,,\,\zeta_1+i\zeta_9\,]\ ,\nonumber
\end{eqnarray}
where  
$$ [\,a\,,\,b\,]~\equiv~a^{[1]}b^{[2]}-a^{[2]}b^{[1]}\ ,$$
is the determinant  of $2$ copies of $a$ and $b$ states. Note that $[\,u_1+iu_2\,,\,\zeta_1+i\zeta_9\,]$ and $[\,\zeta_1+i\zeta_9\,,\,\zeta_8+i\zeta_{16}\,]$ are the highest weights of the $SO(9)$ 
representations $(1001)$ and $(0010)$ respectively. 

\item{$a_1=a_3=a_4=0,\, a_2\ge 1$}

\noindent Here  three copies, $\kappa=1,2,3$ are needed. The $a_2=1$ highest weight solutions  are 

\begin{eqnarray}
&&\theta^1\theta^4\theta^5\theta^8\,[\,u_1+iu_2\,,\,\zeta_1+i\zeta_9\,,\,\zeta_8+i\zeta_{16}\,]\ ,\nonumber\\
&&\theta^1\theta^8[\,u_1+iu_2\,,\,\zeta_1+i\zeta_9\,,\,\zeta_8+i\zeta_{16}\,]\ ,\nonumber\\
&&\theta^1\theta^4\theta^8[\,u_1+iu_2\,,\,\zeta_1+i\zeta_9\,,\,\zeta_8+i\zeta_{16}\,]\ ,\nonumber
\end{eqnarray}
where $[\,u_1+iu_2\,,\zeta_1+i\zeta_9\,,\,\zeta_8+i\zeta_{16}\,]$ is the highest weight of the $SO(9)$ representation $(1010)$, and $[\,a\,,\,b\,,\,c\,]$  is the determinant (antisymmetric product) of $3$ copies of $a,\,b$ and $c$ states.

\item{$a_2=a_3=a_4=0,\, a_1=1$}

\noindent The $F_4$ states are represented by antisymmetric products of $\kappa=2$ copies of $26$ states. The highest weight solutions are

\begin{eqnarray}
&&\theta^1\theta^4\theta^5\theta^8\,([\,u_1+iu_2\,\,,u_3+iu_4\,]+[\,\zeta_1+i\zeta_9\,,\,\zeta_6-i\zeta_{14}\,]+[\,\zeta_8+i\zeta_{16}\,,\,\zeta_3-i\zeta_{11}\,])\ ,\nonumber\\
&&\theta^1\theta^8\,([\,u_1+iu_2\,\,,u_3+iu_4\,]+[\,\zeta_1+i\zeta_9\,,\,\zeta_6-i\zeta_{14}\,]+[\,\zeta_8+i\zeta_{16}\,,\,\zeta_3-i\zeta_{11}\,])\ ,\nonumber\\
&&\theta^1\theta^4\theta^8\,([\,u_1+iu_2\,\,,u_3+iu_4\,]+[\,\zeta_1+i\zeta_9\,,\,\zeta_6-i\zeta_{14}\,]+[\,\zeta_8+i\zeta_{16}\,,\,\zeta_3-i\zeta_{11}\,])\ ,
\nonumber\end{eqnarray}
where $([\,u_1+iu_2\,\,,u_3+iu_4\,]+[\,\zeta_1+i\zeta_9\,,\,\zeta_6-i\zeta_{14}\,]+[\,\zeta_8+i\zeta_{16}\,,\,\zeta_3-i\zeta_{11}\,])$ is the highest weight of the $SO(9)$ representation $(0100)$.
\end{enumerate}
This last case implies that only three copies of $26$ oscillators suffice to generate all $F_4$  representations. It is not possible to construct the $[\,1\,0\,0\,0\,]$ state out of four copies of states in the $\bf 26$. Hence all representations of $F_4$ can be obtained by three copies of harmonic oscillators. 

Since the $T_{ij}$  do not alter the degree of homogeneity of  polynomials in $u_i$ and $\zeta_a$, all solutions are given by the functions $f(u_i,\zeta_a)$ as homogeneous polynomials of their variables. The general highest weight solutions  are  then

\bea
&&\theta^1\theta^4\theta^5\theta^8\,w_1^{a_1}\,w_2^{a_2}\,w_3^{a_3}\,w_4^{a_4}\ ,\nonumber\\
&&\theta^1\theta^8\,w_1^{a_1}\,w_2^{a_2}\,v_3^{ a_3}\,v_4^{ a_4}\ ,\nonumber\\
&&\theta^1\theta^4\theta^8\,w_1^{a_1}\,w_2^{a_2}\,w_3^{a_3}\,v_4^{a_4}\ ,
\nonumber
\eea
where

\bea
w_1&=& [\,u_1+iu_2\,\,,u_3+iu_4\,]+[\,\zeta_1+i\zeta_9\,,\,\zeta_6-i\zeta_{14}\,]+[\,\zeta_8+i\zeta_{16}\,,\,\zeta_3-i\zeta_{11}\,]\ ,\nonumber\\
w_2&=&[\,u_1+iu_2\,,\,\zeta_1+i\zeta_9\,,\,\zeta_8+i\zeta_{16}\,] \ ,\nonumber\\
w_3&=&[\,u_1+iu_2\,,\,\zeta_1+i\zeta_9\,]\ ,\qquad v_3 ~=~[\,\zeta_1+i\zeta_9\,,\,\zeta_8+i\zeta_{16}\,]\ ,\nonumber\\
w_4&=&(u^{}_1+iu^{}_2)\ ,\qquad v_4 ~=~(\zeta^{}_1+i\zeta^{}_9)\nonumber\ .
\eea
All other states are generated by application of the four $SO(9)$ lowering operators.

\subsection{Super Euler Triplets (SET)}
We have just displayed the solutions to Kostant's equation as  products of a $\theta$ part and an internal part that depends polynomially on new variables.  Since the $\theta$ parts describe a superparticle in eleven dimensions, it is tempting to interpret states in the other Euler triplets as superparticles dressed with fields described by these new variables, vector coordinates $u^{[\kappa]}_i$ and twistor coordinates $\zeta^{[\kappa]}_a$. Should we think of these as coordinates (vector and spinor) in the $SO(9)$ space of some other particle? To satisfy the spin-statistics connection, the twistors $\zeta_a$ can only appear quadratically, since odd powers would generate fermions ($SO(9)$ spinors) with Bose properties. This is true of the triplets for which $a_3$ and $a_4$ are even, with no restrictions on $a_1$ and $a_2$.

The lowest Euler triplet which describes supergravity is supersymmetric with an equal number of fermions and bosons. None of the other Euler triplets display space-time supersymmetry, although  a subclass does contain equal number of fermions ($SO(9)$ spinors) and bosons ($SO(9)$ tensors), those for which the Dynkin indices $a_3$ and $a_4$ are even, with no constraints on $a_1$ and $a_2$. Curiously, spin-statistics requires  those super-like Euler triplets (SETs) that share this one feature of supersymmetry, although they are not  space-time supersymmetric by themselves. 

There are four different families of SETs: 

\begin{itemize}

\item The  simplest set has $a_4$   even, and $a_3=a_2=a_1=0$. Their superfields  depend  on  the symmetric products of two vectors  $u^{}_i\,u^{}_j$, and two spinors, $\zeta^{}_a\,\zeta^{}_b$.  The simplest highest weight solutions ($a_4=2$) are  

$$
  \theta^1\theta^4\theta^5\theta^8\,(u^{}_1+iu^{}_2)^2\ ,\qquad\theta^1\theta^8\,(\zeta^{}_1+i\zeta^{}_9)^2\ ,\qquad\theta^1\theta^4\theta^8\,(\zeta^{}_1+i\zeta^{}_9)^2\ ,
$$
where $(u^{}_1+iu^{}_2)^2$ and $(\zeta^{}_1+i\zeta^{}_9)^2$ are the highest weights of $SO(9)$ representations $(2000)$ and $(0002)$, respectively, described by one set of vector and twistor coordinates, which enter as a symmetric second rank tensor represented by $(2000)$ coupled to gravity, and a four-form  $(0002)$, coupled to the three-form and Rarita-Schwinger fields of supergravity.  The highest weight component of its superfield  is given by 

\bean
\lefteqn{\Phi^{}_{(0\,0\,0\,1)}~=~ \phi^{}_{(0\,0\,0\,1)}\theta^1\theta^4\theta^5\theta^8\,(u^{ }_1+iu^{ }_2)^2+}
\\&~~~~~~~+ A^{}_{(0\,0\,0\,1)}\theta^1\theta^8\,(\zeta^{ }_1+i\zeta^{ }_9)^2+\psi^{}_{(0\,0\,0\,1)}\theta^1\theta^4\theta^8\,(\zeta^{ }_1+i\zeta^{ }_9)^2\ ,\eean
where the fields $\phi\ ,\, A\ ,\,\psi$ depend also on the center of mass variables $\zeta^-$ and the transverse vector $\vec u$.  This family requires  one new set of vector and twistor coordinates.

\item The case  $a_3\ne 0$ even, and $a_1=a_2=a_4=0$ requires two sets of extra coordinates, $ \{\,u^{[\kappa]}_i\,\,\zeta_a^{[\kappa]}\}$, $\kappa=1,2$, as its highest weight superfield is  
  
\bean
\lefteqn{\Phi^{}_{(0\,0\,1\,0)}~=~ \phi^{}_{(0\,0\,1\,0)}\theta^1\theta^4\theta^5\theta^8\,[\,u_1+iu_2\,,\,\zeta_1+i\zeta_9\,]^2+}
\\& ~~+ A^{}_{(0\,0\,1\,0)}\theta^1\theta^8\,[\,\zeta_1+i\zeta_9\,,\,\zeta_8+i\zeta_{16}\,]^2+\psi^{}_{(0\,0\,1\,0)}\theta^1\theta^4\theta^8\,[\,u_1+iu_2\,,\,\zeta_1+i\zeta_9\,]^2\ .\eean
The two sets of internal coordinates arrange themselves in the symmetric product of two three forms that couple to the supergravity three-form, and two RS spinors that couples to gravity and the original RS fields of supergravity.  
  
\item When only $a_2\ne 0$, all three supergravity multiplets are dressed by the same $(1010)$ representation, described by triple products of vector and two spinors. The highest weight superfield is now

\bean
\lefteqn{\Phi^{}_{(0\,1\,0\,0)}~=~ \left(\phi^{}_{(0\,1\,0\,0)}\theta^1\theta^4\theta^5\theta^8\,+ A^{}_{(0\,1\,0\,0)}\theta^1\theta^8\, +\psi^{}_{(0\,1\,0\,0)}\theta^1\theta^4\theta^8\right)\,\times}
\\& ~~\times~[\,u_1+iu_2\,,\,\zeta_1+i\zeta_9\,,\,\zeta_8+i\zeta_{16}\,]^{a_2}\ ,\eean
which requires three sets of coordinates $ \{\,u^{[\kappa]}_i\,\,\zeta_a^{[\kappa]}\}$, $\kappa=1,2,3$

\item Finally, if $a_1\ne 0$ only, the three supergravity states are dressed the same way by something with the quantum number of a 2-form, $(0100)$, made up of two vectors and two spinors.
Although it is most complicated in terms of the underlying coordinates, it is simplest in terms of representations. 

\bean
\lefteqn{\Phi^{}_{(1\,0\,0\,0)}~=~\left(\phi^{}_{(1\,0\,0\,0)}\theta^1\theta^4\theta^5\theta^8\, +
 A^{}_{(1\,0\,0\,0)}\theta^1\theta^8\, +\psi^{}_{(1\,0\,0\,0)}\theta^1\theta^4\theta^8\right)
\times} \\
&\times  \left([\,u_1+iu_2\,\,,u_3+iu_4\,]+[\,\zeta_1+i\zeta_9\,,\,\zeta_6-i\zeta_{14}\,]+[\,\zeta_8+i\zeta_{16}\,,\,\zeta_3-i\zeta_{11}\,]\right)\ .\eean
These SETs require  only two copies. This doubling may indicate the presence of $E_6$,  the complex extension of $F_4$. 

\end{itemize}
We note that none of the solutions depend on  the singlet variable $u^{}_0$, which can be traced to the equation  

$$ \Gamma_{}^a\,\frac{\partial}{\partial \zeta_a}~=~0\ .$$
We  may be tempted to  think of these solutions as products of the supergravity ground state and excitations from  new objects (at least two in all but one case)  with both transverse space vector and  spinor (twistor) coordinates. Since vector and spinor coordinates obey Bose statistics, their description is not space-time supersymmetric, although it could be relativistic.

Let us assume that the Euler triplets describe states on which the Poincar\'e symmetries can be implemented. Then one of two possibilities arises:
\begin{itemize}

\item If these states are massless, we already have the necessary ingredients, the $SO(9)$ generators given by the sum $L^{ij}=T^{ij}+S^{ij}$ and there is no further addition to the light-cone Hamiltonian $P_{}^-$. But  these states represent higher spin particles and if they are massless, one can expect grave difficulties in implementing their interactions~\cite{WITTENWEINBERG,DESER}, although there is an infinite number of triplets~\cite{VASILIEV}.   

\item If the excited Euler triplets  describe massive states, we must be able to produce a non-linear realization of  the generators of the massive little group $SO(10)$. This entails the construction of a transverse vector $L^i$, with the commutation relations

$$[\,L^i_{}\, ,\,L^j_{}\,]~=~iM^2_{}\,L^{ij}_{}\ ,$$
where $M^2$ is the mass squared operator which commutes with $L^{ij}$. Then adding $M^2$ to $P^-$, and including $L^i$ in the light-cone boosts satisfies the Poincar\'e algebra. For strings and superstrings, the $L^i$ are cubic in the oscillators, and the commutation relation  works only in the right number of dimensions. 

A necessary condition to realize the massive Euler triplets is to assemble the triplets in $SO(10)$ representations, the massive little group in eleven dimensions, but this is not possible with Euler triplets alone. To see this, consider the $SO(8)$ fermion representations, with Dynkin labels $(1+a_2+a_3,a_1,a_2,1+a_3+a_4)$ with $a_3,a_4$ even. They can be expressed in terms of fields with one spinor index and a tensor structure given by the partition 

$$\{1+a_1+a_2+a_3+\frac{a_3+a_4}{2}\ ,\,a_1+a_2+\frac{a_3+a_4}{2}\ ,\,a_2+\frac{a_3+a_4}{2}\ ,\,\frac{a_3+a_4}{2}\}\ ,$$
and we see that the first row is always larger than the second row. The corresponding $SO(10)$ tensor will also have more in the first row than in its second, but it will also have a partition where the excess in the first row are identified with the tenth direction. This produces an $SO(8)$ tensor with equal number in its first two rows, which is not a triplet spinor. We conclude, in analogy with the Higgs mechanism, that new degrees of freedom are needed to make the Euler triplets massive. If successfull this would describe an object with a massless supersymmetric sector and massive non-supersymmetric states with equal number of fermions and bosons, similar in spirit to Witten's model in $2+1$ dimensions.
\end{itemize}

Either way, our approach still lacks an organizing principle for the inclusion and exclusion of Euler triplets. We have already seen that the spin-statistics connection limits the multiplets ($a_3,a_4$ even), but nothing so far tells us how many Euler triplets should participate. For instance, an infinite number would suggest the inclusion of $F_4$ inside a non-compact group. 

 We see that the Euler triplets generate the spectrum of a Poincar\'e covariant  object which has a ground state with supersymmetry!
This object would be described by its center of mass coordinates $x^-$ and $x_i$, and internal coordinates $u_i^{(\kappa)}\ ,\zeta_a^{(\kappa)}$, where $\kappa$ may run over three values at most. Contrast this with the superstring which is also described by its center of mas coordinates $x^-$, $x_m$, $m=1,\dots, 8$ and an infinite number of internal variables $x^{(n)}_m$, and anticommuting spinor variables $\zeta_\alpha^{(n)}$, $n=1,2,\dots \infty$, with $\alpha=1,\dots 8$. 

In this language, the Euler triplets emerge as much simpler than  superstrings, since they have a finite number of internal variables, although their internal spinor variables satisfy Bose commutations. 

Could these  label the end points of an open  string in the zero tension limit?

Could this new internal space be generated by  the degrees of freedom of the Exceptional Jordan Algebras?

\setcounter{equation}{0}
\section{Exceptional Jordan Algebra}
We have built  Kostant's operator from the coset $F_4/SO(9)$. This coset is a very special projective plane, called the Cayley or Moufang plane, since its  projective geometry  is the only one not to satisfy Desargues' theorem.  Its points can be identified with the projection operators associated with  the quantum mechanical states of the Exceptional Jordan Algebra.  

Jordan algebras~\cite{FEZA} are an alternate description of finite-dimensional Hilbert space in terms of its observables. One defines the symmetric Jordan product

$$
J_a\,\circ\, J_b~=~J_b\,\circ\, J_a\ ,~~~~~~~~~~(\rm I)
$$
which maps observables into observables. It is 
the symmetric matrix product, but since matrices do not commute, the Jordan associator 

$$(\,J_a\,,\,J_b\,,\,J_c\,)~\equiv ~J_a\circ (J_b\circ J_c)- (J_a\circ J_b)\circ J_c\ .
$$
is not necessarily zero, although it satisfies the Jordan identity
 
$$(\,J^{}_a\,,\,J^{}_b\,,\,J^2_a\,)= 0 \ .~~~~~~~~~~(\rm II)
$$
Equations (I) and (II) are the postulates for the commutative but non-associative Jordan Algebras, unlike the matrix multiplication in  Hilbert space which is non-commutative but associative.

To any quantum-mechanical state desribed by the Hilbert space ket $\vert~\alpha >$, corresponds 
the observable  idempotent (density) matrix

$$
P^{}_\alpha={\vert~ \alpha><\alpha~\vert\over<\alpha~\vert~\alpha>}\ ,\qquad 
  P_\alpha\circ P_\alpha =P_\alpha\ .
$$
All familiar quantum mechanics of finite Hilbert space can be expressed in this language.  
For instance, linear dependence among three states translates into
the Jordan statement among their associated projection operators  

$$
\Tr~[(P_\alpha\times P_\beta)\circ P_\gamma]=0 \ ,
$$
using the Freudenthal product 

$$
J_a\,\times\, J_b~\equiv~ J_a\,\circ\, J_b-{1\over 2}J_a\,\Tr(J_b)-{1\over 2}J_b\,\Tr(J_a)
-{1\over 2}\Tr(J_a\circ J_b)+{1\over 2}\Tr(J_a)\,\Tr(J_b)\ .
$$
Unitary maps  in Hilbert space are written  in terms of Jordan operations

$$
\delta P_\alpha~\equiv~ {\cal D}_{B,C}P_\alpha~=~(\,B\,,\,P_\alpha\,,\,C\,)\ ,$$
which reduces to the  commutator relation,  

$$
\delta P_\alpha~=~{1\over 4}[\,[\,B\,,\,C\,]\,,\,P_\alpha\,]\ ,
$$
in complex Hilbert space.  In Jordanese, 
 two observables represented by $A$ and $B$, are compatible if their 
associator $(\,A\,,\,J\,,B\,)$ with any element $J$ of the Jordan algebra  vanishes.

A particular example is time (or light-cone time) evolution   generated by

$$
i\hbar\frac{\partial J}{\partial t}=(\,A\,,\,J\,,\,B\,)\ ,$$
where the Hamiltonian in given by

$$
H={i\over 4}[A,B]\ .$$
Jordan, Von Neumann and Wigner~\cite{JVW} found that the Jordan axioms were also realized  by  $3\times 3$ hermitian matrices over octonions. Octonions, sometimes called Cayley 
numbers, are non-associative generalizations of real, complex and quaternionic numbers. 
This Exceptional Jordan Algebra 
(EJA), has intrigued many people~\cite{FEZA,MURAT,CORE},  but no compelling case for its 
use in physics has ever been made.  The
non-associativity forbids its interpretation  in
terms of 3-dimensional Hilbert space with kets representing physical states. Its  elements  are of the form

$$
J(\alpha_i,\omega_i)~=~\pmatrix{\alpha_1&\omega_3&\overline\omega_2\cr
\overline\omega_3&\alpha_2&\omega_1\cr
\omega_2&\overline\omega_1&\alpha_3}=\sum_1^3\alpha_iE_i+
(\omega_3)_{12}+(\omega_2)_{31}+(\omega_1)_{23}\ ,$$
where $\alpha_i$ are real numbers, and the $\omega_i$ are three 
octonions. Octonions are written in terms of eight real numbers as 

$$ \omega~=~a_0+\sum_1^7a_\alpha~e^{}_\alpha\ ;\qquad \overline 
\omega~=~a_0-\sum_1^7a_\alpha~e^{}_\alpha\ ,$$
where  $e_\alpha$ are the seven imaginary octonion units. 
They satisfy the relations
$$
e^{}_\alpha~e^{}_\beta=-\delta^{}_{\alpha\beta}+\Psi^{}_{\alpha\beta\gamma}~e_\gamma\ ,$$
$\Psi_{\alpha\beta\gamma}$ are  the totally antisymmetric octonion structure 
functions, with only non-zero elements  
$$
\Psi_{123}=\Psi_{246}=\Psi_{435}=\Psi_{651}=\Psi_{572}=\Psi_{714}=\Psi_{367}=1\ .$$
  The Cayley algebra 
is non-associative, but alternative: the associator of three octonions
$$
[\omega_1,\omega_2,\omega_3]~\equiv~ (\omega_1\omega_2)\omega_3-\omega_1(\omega_2\omega_3)
\ ,$$
is completely antisymmetric 
$$
[e^{}_\alpha,e^{}_\beta,e^{}_\gamma]~=~
2\widetilde\Psi_{\alpha\beta\gamma\delta}^{}~e^{}_{\delta}\ ,$$
where
$$
\widetilde\Psi_{\alpha^{}_1\alpha^{}_2\alpha^{}_3\alpha^{}_4}^{}
~=~\frac{1}{3!}\epsilon_{\alpha^{}_1\alpha^{}_2\alpha^{}_3\alpha^{}_4\alpha^{}_5\alpha^{}_6\alpha^{}_7}
\Psi_{\alpha^{}_5\alpha^{}_6\alpha^{}_7}\ ,$$
is the dual of the octonion structure functions.   

Unitary maps on the EJA are replaced by its group of automorphisms, the $52$-parameter exceptional group $F_4$, under which 

$$ F_4:~~~~~~~~\delta J~=~{\cal D}_{h_1,h_2}~J~=~(\,h_1\,,\,J\,,\,h_2\,)\ ,$$
where $h_1,\, h_2$ are $(3\times 3)$ traceless octonionic hermitian matrices,  
each  labelled by $26$ real parameters. Because of the non-associativity of the matrix elements, 
the Jordan associator does not reduce to a commutator.  The traceless Jordan matrices span the $\bf 26$ representations of $F_4$. One can supplement the $F_4$ transformation by an additional $26$ parameters, and define

$$
{\cal D}_X\, J~\equiv~ X\circ J\ ,$$
leading to a group with $78$ parameters. These extra transformations are non-compact, and close on the $F_4$ transformations, leading to the exceptional group $E_{6(-26)}$. The subscript in parenthesis denotes the number of non-compact minus the number of compact generators. 

An important subgroup is $SO(9)$, the automorphism of the $(2\times 2)$ 
Jordan matrices over octonions, generated by transformations that leave an idempotent, say $E_3$,  invariant  
 
$$
{\cal D}_{(\omega)_{12},(\tau)_{12}}\ ,~{\cal D}_{(\omega)_{12},E_1-E_2}\ ,$$
where $\omega$ and $\tau$ are octonions. The first, antisymmetric under $\omega\leftrightarrow \tau$, represents the $28$ transformations of $SO(8)$. Traceless $(2\times 2)$ matrices transform as the nine components of the vector representation of $SO(9)$.

It follows that the EJA dynamic evolution is generated by  $F_4$ transformations, which can be catalogued in terms of unbroken symmetries.  Could  the $SO(9)$ subgroup of EJA automorphisms    be identified with the light-cone little group in eleven space-time dimensions~\cite{PR2}? 

EJA states are represented by points in the projective geometry over $F_4/SO(9)$, which  can be written in the form

$$
P=\Omega\frac{1}{\sqrt{\Omega^\dagger\Omega}}\Omega^\dagger_{}\ ,$$
where $\Omega^T= (\omega_1,\omega_2, \omega_3)$ is an octonionic vector. With
seven redundant phases and one normalization condition, one of the three octonions  can be set equal to one, leaving us  with $16$ 
real parameters to determine a point, the labels of the Moufang  projective space. It is also   the coset space  $F_4/SO(9)$, since  $F_4$ acts on the points, while  its 
$SO(9)$ subgroup leaves any one point invariant, and the 
transformations in $F_4/SO(9)$ map this point into other points.
As mentioned earlier, this projective geometry is unique as it does  not satisfy Desargues' theorem. 

For those who do not remember that theorem: take three lines $p,q,r$ that meet at a
point. Take any two points on each of these lines,  call them $p_1,p_2$,
$q_1,q_2$ and $r_1,r_2$. The lines connecting $p_1q_1$ and $p_2q_2$
meet at the point $A_{pq}$, while $p_1r_1$ and $p_2r_2$ meet at
$A_{pr}$. Finally  the lines  $q_1r_1$ and $q_2r_2$ meet at $A_{qr}$. 
Desargues theorem states that the three points $A_{pq},A_{pr},A_{qr}$ 
lie on the same line.  

 To conclude, we have investigated several schemes which  involve exceptional groups in the description of space(-time); especially interesting is the connection between the light-cone little group in eleven dimensions and $F_4$. In the process we have shown how the superparticle could be generalized to include Euler triplets. Also we have established some curious mathematical links between   EJA quantum mechanics~\cite{MURAT2} and Euler triplets. It   behooves us to continue to investigate~\cite{NEXT} these relations and endow them with physical significance. 

\vspace{5 mm}
\noindent {\bf Acknowledgements}:
I wish to congratulate  Marc Henneaux for the recognition he so richly deserves, and to thank him for his kind invitation to this exceptional meeting, as well as Dr. L. Eyckmans and the Francqui Foundation for hospitality {\it sans pareil}. This work is  done in collaboration with Lars Brink and X. Xiong. I also wish to thank A. Khan for discussions.   
The author is also supported in part
by the US Department of Energy under grant DE-FG02-97ER41029.

%\section {References}

\end{document}